# Kinetic inductance driven nanoscale 2D and 3D THz transmission lines

S. Hossein Mousavi[1], Ian A. D. Williamson[1], Zheng Wang[1,a)]

[1]Microelectronics Research Center, Department of Electrical and Computer Engineering, The University of Texas at Austin, Austin, TX 78758 US

Abstract

We examine the unusual dispersion and attenuation of transverse electromagnetic waves in the few-THz regime on nanoscale graphene and copper transmission lines. Conventionally, such propagation has been considered to be highly dispersive, due to the RC-constant-driven voltage diffusion below 1THz and plasmonic effects at higher frequencies. Our numerical modeling between the microwave and optical regimes reveals that conductor kinetic inductance creates an ultra-broadband LC region. This resultant frequency-independent attenuation is an ideal characteristic that is known to be non-existent in macro-scale transmission lines. The kinetic-LC frequency range is dictated by the structural dimensionality and the free-carrier scattering rate of the conductor material. Moreover, up to 40x wavelength reduction is observed in graphene transmission lines.

Transmission lines, one of the simplest forms of single-mode electromagnetic waveguides, are widely used for signaling in digital, analog, and optoelectronic systems. Transmission lines generally consist of two or more parallel conductors that guide a transverse electromagnetic (TEM) wave over many wavelengths. TEM waveguides are unique in that they lack a low-frequency cut-off. However, at higher frequencies, they become band-limited by an attenuation that increases with frequency, caused by the skin effect and the closely-related proximity effect.[1] Ideal transmission lines should exhibit constant group delay and flat-attenuation over a broad bandwidth. Such an ideal characteristic occurs only in the so-called LC region, in which conductor inductance dominates over resistance. Unfortunately, the lateral dimensions of practical transmission lines results in a negligible bandwidth for the LC regime[1], as the onset frequency of the LC regime nearly coincides with the onset of the skin effect. The skin effect imparts a square-root frequency dependence on the attenuation[2], imposing a fundamental performance bottleneck. Practically, the skin effect limits the maximum length of board- and chip-level interconnects[3,4], the bandwidth of traveling-wave modulators[5–8], and the spatial resolution of time-domain reflectometry systems[9].

Two recent advancements have made it important to re-examine the performance regions of nanoscale transmission line: state-of-art fabrication processes have allowed the dimension of transmission lines to fall below the skin depth; graphene has emerged as a practical conductor with large kinetic inductance[10–15]. Both can be exploited to suppress the skin effect and subsequently realize frequency-independent attenuation. The kinetic inductance stems from the kinetic energy of oscillating free charges, and is significant in superconductors[16] at microwave frequencies and metals at optical frequencies[17]. Graphene is in that is has a dominant kinetic inductance at relatively low THz frequencies[18]. This high kinetic inductance has been exploited for ultra-short-wavelength graphene plasmons for infinite sheets[18–20] and nano-ribbons[21–23]. Although various transmission line structures have been studied for graphene, separately in the microwave[21,24,25] and optical regimes[20,22], to date no comprehensive study has been conducted to map the various performance regions (RC, LC, skin effect, and plasmonics regimes) of graphene transmission lines and their shared underlying physical mechanisms over a wide range of graphene sizes and operating frequency ranges.

In this letter, we carry out an ab-initio study of the effect of material kinetic inductance on the propagation and attenuation of the transverse electromagnetic modes in nanoscale and microscale transmission lines, and its potential in improving performance for signaling and sensing applications. Over a vast space spanning $10^6$–fold variation in dimensions (nm-through-cm) and $10^4$–fold variation in frequency (10GHz-100THz), we examine the conditions for the kinetic inductance to produce a surprisingly broadband LC region for small-scale transmission lines. Investigating both graphene and copper conductors, we develop a map of the performance regions for 2D and 3D materials, contrast the significant differences in their scaling laws and explain dominant physical mechanisms that limit their performance.

Among various transmission line structures, we focus on the coplanar stripline (CPS) shown in Fig 1(a) because it exhibits the typical skin-effect induced attenuation and it has been widely used in integrated systems[26] and traveling-wave modulators[7]. The CPS consists of two parallel conductors on top of a dielectric substrate[27,28]. We consider two distinct CPS designs with two different conductor materials: one with single-layer graphene ribbons and the second with copper wires. The copper wires have a finite thickness, $t$, and the graphene ribbons are treated as an infinitesimally-thin surface conductor. At THz frequencies and below, both materials are accurately modeled with a Drude conductivity[19],

$$\sigma(\omega) = \frac{\sigma_0}{1 + j\omega/\Gamma} \quad (1)$$

where $\Gamma$ is the phenomenological scattering (collision) rate of the free carriers[18], encompassing all scattering mechanisms. $\sigma_0$ corresponds to the bulk DC conductivity for copper and surface conductivity for graphene. Scattering due to surface roughness, finite temperature effects and spatial dispersion are neglected. The effective surface conductivity (i.e., $\sigma_s = \sigma \times t$) of a 3 nm-thick copper film is roughly 10x larger than the graphene surface conductivity (Fig. 1b). Thicker copper films are used in the rest of this paper, implying an even higher conductivity for copper lines. The scattering rate is taken as 1 THz for graphene and as 7.26 THz[29] for copper.

The underlying physical mechanisms leading to frequency-dependent attenuation in transmission line modes are captured by the circuit parameters in the solution to the telegrapher's equations. Since we limit our discussion to lateral dimensions much smaller than free-space wavelength, this analytical model is accurate and yields comparable results to ab-initio solution to the Maxwell equations.[1,17] The real and imaginary parts of the complex-valued propagation constant[30], $k = \alpha + j\beta = \sqrt{j\omega C (R + j\omega L_F + j\omega L_K)}$, reflects the spatial phase evolution ($\beta$) and attenuation ($\alpha$). The circuit parameters $C$, $R$, $L_F$, and $L_K$ represent the inter-conductor capacitance, conductor resistance, Faraday inductance, and kinetic inductance per unit length, respectively. The LC region occurs at high frequencies, where $\omega(L_F + L_k)$ exceeds $R$, but at low enough frequencies where $R$, $L_F$, and $L_K$ remain constant. Unlike the highly dispersive RC region at lower frequencies, the LC region exhibits linear group velocity and constant attenuation. However, at high frequencies, $R$ and $L_F$ usually vary with frequency due to the non-uniform conduction current distribution. In particular, a large part of $L_F$ comes from the magnetic flux internal to the conductors, which can be greatly reduced by current crowding to the conductor surface. As the current distributes itself minimize the total impedance $R + j\omega L_F$, current crowding occurs quickly above the LC on-set frequency to reduce $j\omega L_F$, at a cost of increased and frequency-dependent resistance and attenuation[1]. In other words, without the kinetic inductance, the onset frequency of the skin effect and the LC region are very close, because the magnetic flux within the conductors plays a major role in the total impedance. In order to obtain a large LC bandwidth without kinetic inductance, a large conductor separation would be needed to increase the ratio of the external to internal inductance.

In contrast to the Faraday inductance, the kinetic inductance stems from a material constant, the complex-valued conductivity. The kinetic inductance impedance is proportional to the imaginary part of the reciprocal of the conductivity, while the well-known resistive impedance relates to the real part. Physically, they contribute to the power dissipated or stored, respectively, in the conduction currents: $J^* \cdot E = \sigma^{-1} |J|^2 \propto RI^2 + j\omega L_K I^2$. $J$ is the surface or volume current density produced by an electric field $E$. It is worth noting that the ratio of the kinetic inductive impedance to the resistance, $\omega L_K / R = \text{Im}\{\sigma^{-1}\} / \text{Re}\{\sigma^{-1}\}$, is a material constant that is completely independent of the geometry. Thus, similarly to resistance, the kinetic inductance is *inversely* proportional to the current carrying area $A$ (or width $w$ in the case of graphene), as $L_K = \text{Im}\{\sigma^{-1}\} \omega^{-1} A^{-1}$. Assuming a Drude model, both impedances become frequency independent, even at frequencies above the scattering rate, as $L_K = \sigma_0^{-1} A^{-1} \Gamma^{-1}$ and $R = \Gamma L_K$. In contrast to the Faraday inductance, the kinetic inductance *increases* with current crowding, and thus suppresses its occurrence after the LC onset in transmission lines that have a dominant kinetic inductance.

In nanoscale transmission lines, the kinetic inductance can significantly exceed the Faraday inductance. Generally, the Faraday inductance of a transmission line is scale-invariant, $L_F = L_0 \ln(w/g)$, where $L_0$ is a geometry dependent constant[3]. Given a specific aspect ratio for the conductor size and separation (that is, $t/w = m$ and $g/w = m'$ with $m$ and $m'$ being constant), and scaling $w$ from nm to cm, the size threshold at which the kinetic inductance exceeds the Faraday inductance occurs, at $w_{K,3D} = 1/\sqrt{\sigma_0 \Gamma m L_0 \ln(1/m')}$ for 3D conductors and at $w_{K,2D} = 1/[\sigma_0 \Gamma L_0 \ln(1/m')]$ for 2D conductors. For the parameters chosen in this paper, $w_{K,3D}$ and $w_{K,2D}$ are on the order of 100 nm and 100 μm respectively. In contrast to 2D material lines, the threshold in 3D material lines is much smaller. This is partly due to the square root dependence and partly due to their larger DC conductivity (Fig. 1b). If a lower material scattering rates can be achieved, the threshold can be increased, with more mileage being gained in 2D lines.

Below such size thresholds, the onset of LC region ($\omega_{LC}L = R$) reduces to $\omega_{LC} = R/L_K = \Gamma$. This surprisingly simple relationship suggests that LC region and the associated constant attenuation can be found immediately above the intrinsic scattering rate (a constant for a given material). This behavior stands in contrast to transmission lines dominated by the Faraday inductance, in which $\omega_{LC}$ is strongly dependent on the conductor size and the geometry. Therefore, two design conditions, $w < w_K$ and $\omega > \Gamma$, must be both met to achieve a broad LC response.

Finite-element ab-initio modeling of Maxwell's equations corroborates this observation, as we extract inductive impedances and the resistance from the calculated fields (Fig.2). We consider a 25x dimensional scaling of a copper CPS (Fig.2a) and a graphene CPS (Fig.2b), from $w$ = 20 nm to 500 nm in a regime with large kinetic inductance. The conductor separation (as well as the copper conductor thickness) are equal to $w/2$ (i.e. $m = m' = 1/2$). The resistance and the kinetic inductance both decrease by 25x for the graphene CPS and by 625x for the copper CPS throughout the spectral range. In contrast, the Faraday inductive impedance (grey curves in Fig.2) is unaffected by the scaling. The difference between the 2D and 3D systems lies in the fact that the kinetic inductance completely dominates the Faraday inductance in the graphene CPS, while the copper CPS needs to be reduced to tens of nanometers in width before the kinetic inductance becomes dominant. The resultant LC onset frequency, $\omega_{LC}$, i.e. the cross-over between the resistance and the *total* inductive impedance, exhibits different scale-dependence for 2D and 3D transmission lines at nanoscale. For a graphene CPS, with a dominant kinetic inductance, its onset frequency is unchanged by scaling, indicated by the circles in Fig. 2b. For a copper CPS with a width of 500nm, the Faraday inductance dominates. Although $R$ and $\omega L_K$ intersect at the same frequency as graphene, $\omega_{LC}$ is instead determined by the intersection of $R$ and $\omega L_F$ at 1.5THz. The dominance of the Faraday inductance in copper is rather a consequence of much reduced kinetic inductance, due to a higher DC conductivity from the larger cross section of a 3D CPS and a large scattering rate $\Gamma$ for copper. These two factors associate the kinetic-LC region with copper transmission lines narrower than 50 nm, a much smaller size than that in graphene.

These observations are supported by numerically-computed spectra of real and imaginary parts of *k*, i.e. the phase propagation and attenuation, for several copper and graphene CPS's in Fig. 3a and 3b. The LC onset corresponds to a divergence of the real and imaginary components from their shared values in the RC region. The LC region manifests itself as an adjacent frequency range with a constant real part of *k*. For graphene (Fig. 3a), regardless of its width, the LC region starts exactly at the fixed intersection frequency of $\omega L_K$ and R seen in Fig 2(a). Even for the widest graphene CPS considered at *w* = 500 nm, the kinetic inductance dominates. The copper CPS's, on the other hand, have an LC onset that shifts to lower values for larger electrode sizes (Fig. 3b), as expected for a transmission line dominated by Faraday inductance. Among the four widths, only the narrowest case of 50 nm exhibits a dominant kinetic inductance.

With the LC region in the THz regime for nanoscale transmission lines, the upper frequency limit of the LC region is determined by a set of mechanisms qualitatively different from that in microscale and macroscale transmission lines. In copper and other metals, for frequencies below their intrinsic scattering rate $\Gamma$ around 7.26 THz, the skin depth is collisional and scales as $1/\sqrt{\omega}$, whereas above $\Gamma$, the skin depth converges to a fixed value around 100nm[17]. Thus, no skin effect is seen for copper CPS's which are narrower than 100nm (Fig. 3b), and the kinetic LC region extends and blends into the plasmonic regime at optical frequencies. For graphene, however, the upper frequency limit for the LC region is not defined by the skin effect, but rather by the self-capacitance between lateral parts of a graphene ribbon. As $\beta(\omega)$ of the TEM modes approaches the graphene plasmon dispersion relation (Fig. 3a), the self-capacitance introduces significant lateral fields and currents, similar to the transition between metal optics and plasmonics in metallic nanostructures[17]. The resultant redistribution of current towards the edges of the ribbon (insets of Figure 3c), and the added lateral current, create a capacitive "edge effect" that increases attenuation sharply beyond 30 THz.

For graphene (Fig 3c) and copper (Fig 3d), the overall landscape of TEM wave propagation and attenuation is defined by the boundaries separating the previously unrecognized kinetic-LC region (green), the conventional RC region[11] (pink), the skin/edge effect region (yellow), and the plasmonic region[20] (blue). All of the transmission line dimensions are proportionally scaled with respect to the electrode width from tens of nanometers to one centimeter.

The lower bound of the LC region (the dark red curve), $\omega_{LC} = R/(L_F + L_K)$, manifests itself as a size-independent horizontal line at nanoscales, where the kinetic inductance dominates. When the width exceeds 100 nm for copper and 100 μm for graphene, the Faraday inductance becomes dominant, and thus the LC onset occurs at a width-dependent lower frequency. The slope of this transition is a function of the material dimensionality, a steeper $w^{-2}$ dependence for the 3D copper CPS, and a more gradual $w^{-1}$ dependence for the graphene CPS.

The upper bound of the LC region (the dark green curve) is defined by the onset of the skin effect region (or edge effect region for nanoscale graphene). For copper, this boundary is defined by the skin depth $\delta$ becoming half of the electrode width, $w$. It follows the same well-recognized $w^{-2}$ dependence as the LC onset. This boundary becomes a vertical asymptote above the intrinsic scattering rate, as the skin depth converges to its constant collision-less value[17]. The overall behavior of the copper CPS is predominantly characterized by the skin effect above 200nm in width as predicted by conventional wisdom, and by a surprisingly wide kinetic LC region above 8 THz. For single-layer graphene, although it has an infinitesimal thickness and is free from skin effect in the conventional sense, its LC region does transition into an edge effect region at a frequency empirically chosen such that the calculated line resistance exceeds that of a spatially uniform current by 5%. Above the plasmon cut-off frequency (horizontal branch of the dark blue curves, see supplemental info) of an infinitely-wide graphene sheet, this boundary has a dependence of approximately $w^{-0.5}$. And below this plasmon cut-off frequency, the LC-skin effect boundary is close to the RC-LC boundary, suggesting that graphene ribbons can exhibit current crowding near the edges when their width exceeds 100 μm.

The third boundary (the dark blue curve) separate the skin/edge effect region from the plasmonic region. In the plasmonic region, each conductor can support its own surface plasmon, and the transmission line becomes multimoded. For graphene, this boundary occurs when the wavelength of an infinitely-wide-graphene plasmon falls below twice the ribbon width, and runs parallel to the LC-edge-effect boundary, suggesting that the observed edge effect is associated with the transition from quasi-TEM modes to ribbon plasmons. On the other hand, for copper transmission lines, the plasmonic region emerges around the scattering rate of copper, and ends when the conductor width falls below the collision-less skin depth (Fig. 3d). The skin-effect region vanishes for conductor widths that are below 100 nm.

Contrasting graphene transmission lines to their copper counterparts, we observe that graphene can provide a much lower onset frequency for the kinetic LC region. Its LC region allows wider structures up to 100 µm, and features a more gradual frequency slope beyond. However, when absolute lowest attenuation is desired and dispersive attenuation can be tolerated, copper (and other metal) transmission lines are more preferable. Multi-layer graphene can also be used[31], but is expected to exhibit performance similar to 3D copper. Finally, for conductor sizes below 100 nm, copper lines suffer from reduced conductance due to scattering at grain boundaries and surface roughness[32], which has been partially accounted for in this paper with a reduced DC conductivity for copper at $2 \times 10^7 S \cdot m$. Surface conductivity of graphene nanoribbons also deviate from the Drude model considered in this paper, due to the large contribution from conducting edge modes[33].

It is worth noting that in comparison to graphene plasmons known for their exceptional wavelength reduction[18,22], the TEM modes of a graphene transmission line exhibit even greater reduction at few-THz frequencies (Fig. 3a). The dispersion relation of a graphene plasmon[19] suggests a low-frequency radiative cutoff (see supplementary), e.g. 1.5THz for the graphene considered here. The TEM mode on the other hand, remains guided throughout the entire spectral range, and also has a much smaller wavelength. Additionally, finite-width graphene ribbons support plasmons exhibiting even larger $\beta$ than those of the infinitely-wide graphene films, but experience even higher cut-off frequencies. Thus, in the spectral range between 1 and 10THz, the TEM mode provides the optimal combination of wavelength-reduction and constant attenuation.

In conclusion, first-principle modeling of transmission lines has revealed that the well-known TEM modes enjoy a previously unnoticed broadband constant-attenuation regime, when the lateral dimensions are reduced to the nanoscale for both 2D and 3D conductors. This so-called kinetic-LC regime is induced by a predominant kinetic inductive impedance. The threshold dimension for this regime in a 3D line is defined by the collision-less skin depth of the conductor, about 100 nm for copper. Graphene transmission lines are remarkable in that their kinetic-LC region extends to relatively large micron-scale widths. The onset frequency of the kinetic-LC region is fixed at the free-carrier scattering rate, a material property, and therefore high-quality graphene with a low scattering rate would be desirable for THz/sub-THz applications. The upper frequency cutoff, on the other hand, depends on the material dimensionality: 2D graphene lines exhibit a width-dependent upper cutoff induced by the plasmonic "edge-effect", whereas 3D lines have no upper cutoff in the spectral range in which the Drude model is valid. Although graphene lines exhibit constant attenuation at larger dimensions, copper lines of comparable widths provide the absolute lowest attenuation. However, graphene's large kinetic inductance drastically reduces the mode wavelength by 10-40 fold, approaching 1 micron at 10THz. This deep subwavelength operation and the inherent tunability of graphene open up new opportunities for conventional signaling and sensing applications, as well as emerging applications involving metamaterials.

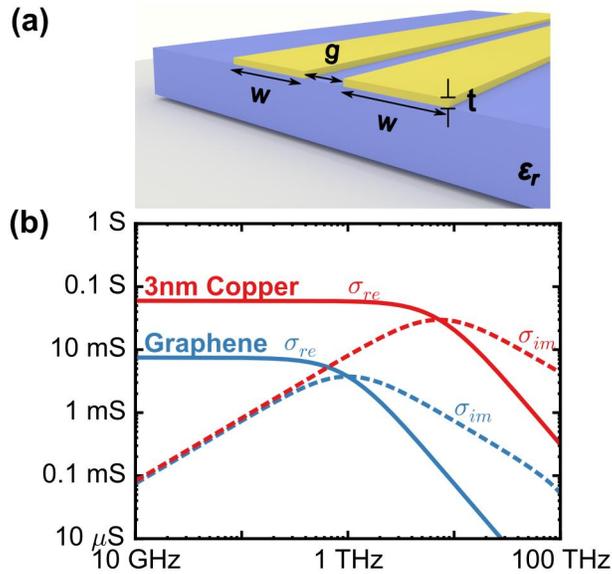

Figure 1: (a) Schematic of a coplanar stripline consisting of two parallel conductors on a $SiO_2$ substrate ( $\epsilon_r = 2.25$ ) (b) Complex-valued surface conductivity of single-layer graphene (blue) compared with a 3 nm copper film (red). Graphene has a Fermi level of 0.4 eV and a phenomenological scattering rate of 1 THz and copper has a phenomenological scattering rate of 7.26 THz.

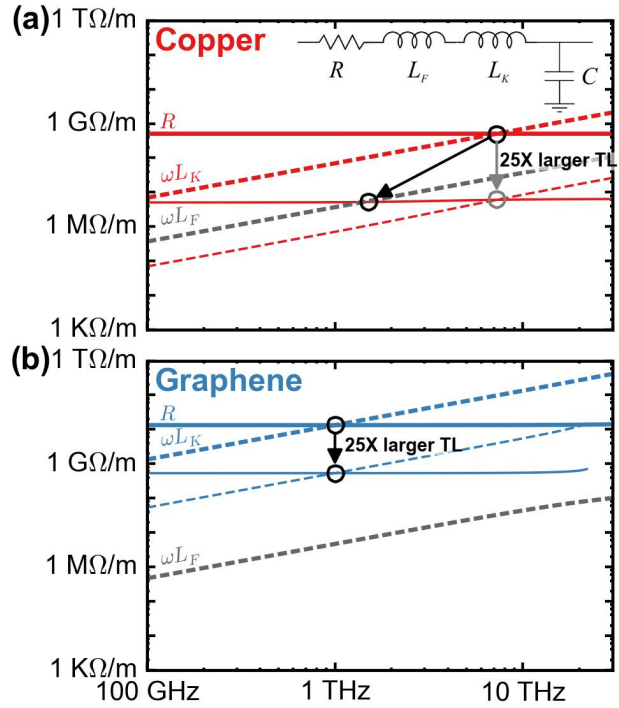

Figure 2: Effect of geometric scaling on the distributed circuit parameters of coplanar striplines made of (a) copper and (b) graphene. A 25x size increase, from (w=20 nm, g=10 nm) to (w=500 nm, g=250 nm) (and additionally t=10 nm to 250 nm for copper), reduces the resistance R (colored solid curves) and the kinetic inductance component $\omega L_k$ (dashed curves) by 625x for copper and 25x for graphene. The Faraday inductive impedance $\omega L_F$ (gray curves) remains unchanged under scaling in both cases, due to a fixed w/g ratio.

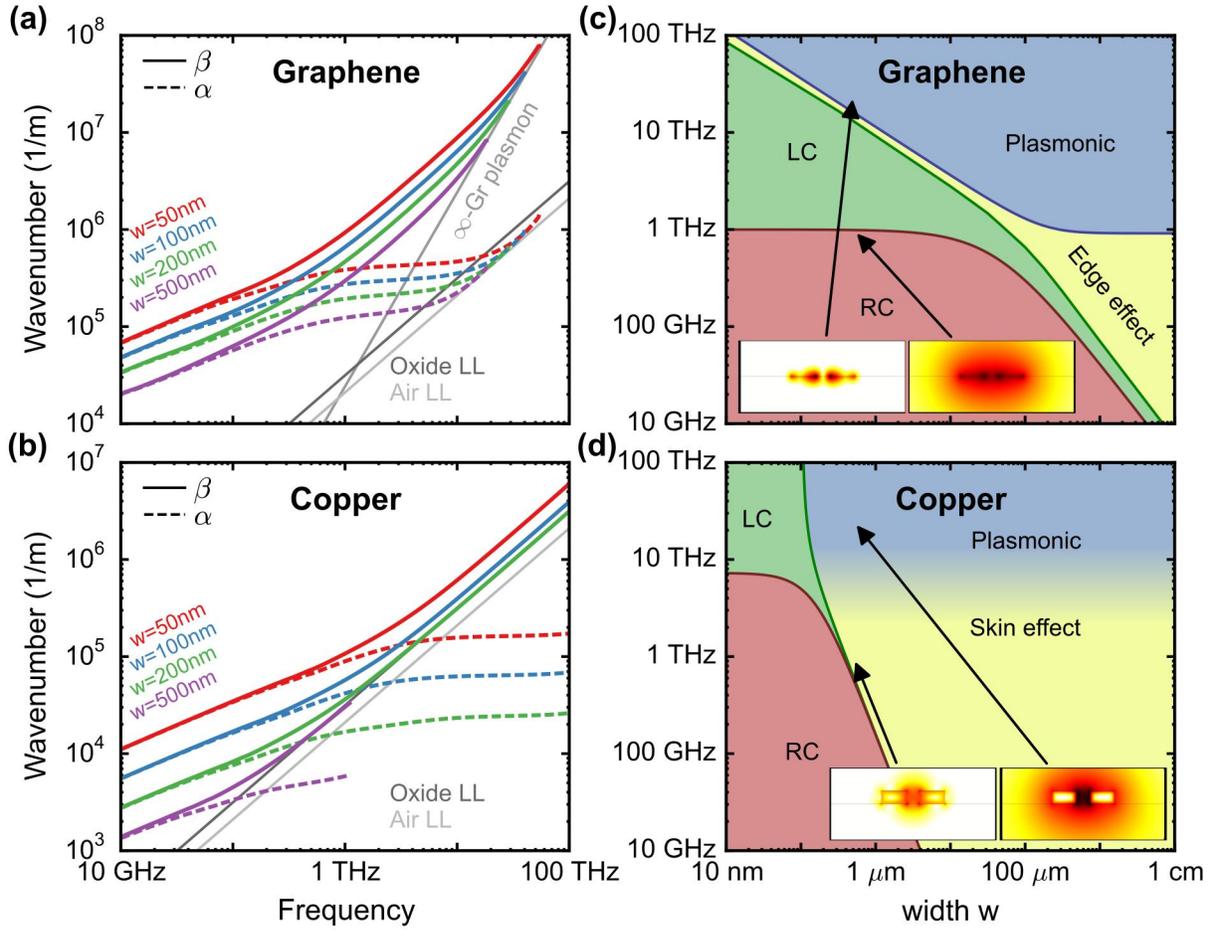

Figure 3: (a) Attenuation and dispersion spectra for graphene coplanar striplines with various conductor widths. (b) Performance regimes of a graphene coplanar stripline as it scales from nanoscale to macroscale. (c) Attenuation and dispersion spectra for copper coplanar striplines with various conductor width. (d) Performance regimes for a copper coplanar stripline. $g$ and $t$ (for copper) are proportionally scaled as $g = w/2$ and $t = w/2$. Lower boundary of skin effect regime is $\delta_{skin}(\omega) = w$. Top boundary of RC regime is $\omega = R/(L_k + L_f)$. Lower boundary of plasmonic/multimode regime corresponds to $w = \lambda_{spp}/2$. Insets illustrate the magnetic energy density in log scale.